# User Popularity-based Packet Scheduling for Congestion Control in Ad-hoc Social Networks

Feng Xia[1], Hannan Bin Liaqat[1], Ahmedin Mohammed Ahmed[1], Li Liu[1], Jianhua Ma[2], Runhe Huang[2], Amr Tolba[3,4]

[1] *School of Software, Dalian University of Technology, Dalian 116620, China*
[2] *Faculty of Computer and Information Sciences, Hosei University, Japan*
[3] *Riyadh Community College, King Saud University, Riyadh 11437, Saudi Arabia*
[4] *Mathematics and Computer Science Department, Faculty of Science, Menoufia University, Egypt*

A B S T R A C T

Traditional ad-hoc network packet scheduling schemes cannot fulfill the requirements of proximity-based ad-hoc social networks (ASNETs) and they do not behave properly in congested environments. To address this issue, we propose a user popularity-based packet scheduling scheme for congestion control in ASNETs called Pop-aware. The proposed algorithm exploits social popularity of sender nodes to prioritize all incoming flows. Pop-aware also provides fairness of service received by each flow. We evaluate the performance of Pop-aware through a series of simulations. In comparison with some existing scheduling algorithms, Pop-aware performs better in terms of control overhead, total overhead, average throughput, packet loss rate, packet delivery rate and average delay.

## 1. Introduction

Mobile Social Networking in Proximity (MSNP) is based on the principle of geo proximity and social relationship which allows users to know about the availability of nodes in social networks [1]. The nodes that are involved in Mobile Social Networks (MSNs) use intrinsic content features with context of items and the objectivity of human for better searching of relevant users [2]. In MSN, the MSNP provides direct connectivity to users when they are in proximity. This technique is helpful to reduce the user's data cost which solves the issue of unnecessary data traffic, passing through Internet. An Ad-hoc Social Networks (ASNETs) is a branch of MSNP that has emerged as a new area of research with multiple innovative applications [3]. It facilitates users to communicate with each other using social features such as similarity, centrality, community, social graph, human mobility pattern and tie strength [4]. The nodes in ASNETs communicate with each other using scarce wireless interface and node capacity. Accordingly, if there is a huge number of data transfer on a single intermediate node without any proper scheduling mechanism then ASNETs involves in congestion. The occurrence of congestion in ASNETs is due to the reason of some popular sender nodes when they come in proximity area and transfer large amount of data. Thus, if any type of delay occurs in transferring data packets of popular node on intermediate node then congestion occurs. Although, the throughput of ASNETs and reduction of congestion on the intermediate node can be improved by designing a packet scheduling algorithm that works according to the concept of ASNETs or MSNP. This is because packet scheduling provides efficient sharing of bandwidth resources among nodes. It also provides solution to the problems caused by multiple connections after sharing one link.

To improve the performance of scheduling algorithm, existing research provides an efficient solution at Medium Access Control (MAC) layer (such as reduction in collision at the MAC layer and fairness in bandwidth sharing [5]-[7]) and routing layer [8]-[10]. However, the work related to efficient queue scheduling mechanisms for wireless interface has not been broadly addressed specially in ASNETs environments. This occurs because ASNETs provides social property (Popularity) based communications between nodes, whereas existing ad-hoc networks use simple priority scheduling such as First-In-First-Out (FIFO) order. Scheduling the data packets based on popularity means that we forward the data packet of the most popular node initially from the intermediate node, which can solve the congestion related problem [11]. This is due to most popular node receiving lots of data from other nodes that create congestion in the intermediate node at an earlier stage. In another aspect, the availability of the most popular node signifies that it has a strong social strength and importance in the network. Consequently, for avoiding the dropping of most popular node's data packets from an intermediate node, the earlier transfer of it is necessary. Scheduling based on social popularity also plays a vital role in increasing the system throughput. As a result of the above reasons, we need to solve such congestion issues by utilizing maximum bandwidth without wastage of resources. Similarly, our previous work (TIBIAS), [12] also solved the congestion related issues in ASNETs and provided maximum utilization of resources after adjusting the data rate with proper bandwidth sharing scheme. TIBIAS provided a sender side solution, which was based on the transport layer and it used social property (*similarity*) for setting the data rate. As compared to TIBIAS, this scheme provides intermediate based solution that manages queue at network layer. The management of queue is based on the proper scheduling of the data packets. Furthermore, for efficient utilization of resources, Pop-aware employs the degree centrality social property.

----------

* Corresponding author: Feng Xia; email: f.xia@ieee.org; tel: +86-411-62274391

In order to solve above scheduling issues, this paper proposes a novel data-scheduling algorithm called User Popularity-based Packet Scheduling for Congestion Control in Ad-hoc Social Networks (Pop-aware). Our algorithm is divided into two main processes. First, to make proper decisions for scheduling, it calculates the load of data packet at intermediate node. After calculating it, we start the scheduling scheme. Secondly, we set prioritization value to the flow. To provide the efficient solution in an MSNP based ASNET model, our Pop-aware algorithm sets prioritization based on the *degree centrality* social property, which indicates the popularity level of a node's data packets.

In this paper, we present and discuss an extended version of our work [13]. We provide a broader view of literature in terms of related work. This paper also enhances the algorithm description (pseudocode) by involving cases regarding the introduction of new data when any node is proximal to intermediate node. Furthermore, in comparison to [13], this paper provide theoretical analysis of scheme and conducts further simulations using appropriate evaluation metrics such as control overhead, total overhead, packet delivery rate and average delay with number of connections. In order to briefly understand the concept of degree centrality, we also present mathematical calculations. This is done to further assess and ascertain the performance of our proposed algorithm. In summary, the contribution of this work includes:

- **Avoiding the packet dropping problem:** Pop-aware calculates the load at the intermediate node to start the scheduling process. This decision helps to avoid the dropping of the most popular node's data.
- **Controlling congestion effectively:** Pop-aware provides a higher priority to the flow of data packets that have highest degree centrality. It is helpful to control the congestion at the intermediate node and fully utilize the available bandwidth.
- **Fairness:** We devise a mechanism that can provide fairness of service received for each flow based on served and non-served concept. It can also provide fair utilization of resources between nodes after calculating the throughput ratio for each flow.
- **Proper scheduling decision:** We propose a set of scheduling techniques on the arrival of new flow data packets after calculating its rates and *degree centrality* values to reduce delivery delay, while achieving higher throughput.
- **Extensive simulations:** We conducted comprehensive simulations to evaluate Pop-aware's performance in comparison with other scheduling protocols.

The rest of the paper is organized as follows: related work is discussed in the next section. Section 3 describes network model and preliminaries. Section 4 presents our Pop-aware scheduling algorithm with design problems. In Section 5, we present the performance analysis and further calculation of methodology with discussions of evaluations is provided in Section 6. Finally, we conclude the paper in Section 7.

## 2. Related work

In MSNs, stream data and the life logs use large amount of internet resources and their combination is called social streams. This social stream utilizes large amount of internet resources but provides easiness in collection of information and social activities of users [14]. The online MSNs exploit social behaviors among users for better utilization of resources. These social behaviors depend upon the social interaction and connectivity among users that are helpful to enhance the performance of system [15] [16]. As compared to online MSN, MSNP reduces the cost of communication after using ASNET concept. In ASNETs, communication between nodes is without any infrastructure. Therefore, in ASNETs nodes are assigned to different groups that are based on semantic matching of users' interest [17]. Furthermore, users exploit some social properties such as centrality, social relationship and community to work in an ASNETs. In MSNs, congestion is the basic reason that creates hurdle in communication. The reason for the occurrence of congestion can also be the non-proper scheduling of the data packet in the network. However, this congestion issue can create problems for prioritized flow after dropping data packet of prioritized node from the intermediate node. Consequently, to solve this issue it is necessary to serve the most prioritized node's data packet first, with fair resource allocation and controlling of congestion [18]. To provide the scheduling solution in ASNETs, we need to consider social properties for communication between nodes. Thus, in this paper we use *degree centrality* social property for communication between nodes and schedule data packets accordingly. In the remaining part of this section, we discuss the existing work in term of *centrality* (social property) and elaborate the above issues. Moreover, we review literature in line of MSNs and ad-hoc networks.

## 2.1. Centrality based congestion control in mobile social networks

In social opportunistic networks, where node-to-node or human-to-human communication exists, unfair distribution of resource allocation can create a major problem. The nodes that are involved in opportunistic networks behave intermittently [19]. Moreover, the operations of some devices, involved in this network are also based on human movement as humans are the main carriers of mobile devices. The dynamic and unpredictable movements of nodes increase the delay of data transformation that also affects the cost of transportation [20]. To elaborate social activity of nodes, this subsection discusses existing literature work on *degree centrality, betweeness centrality, closeness centrality* and *ego betweeness centrality*. The centrality based communication shows that these social properties are helpful in traffic distribution and management of data storage to avoid the congestion. Furthermore, to divide large networks into clusters the *edge centrality* provides high advantage [21].

To provide the structure of MSNs, Hossmann *et al.* [22] proposed a scheme where they show that the structures of MSNs are non-random. This non-random behavior illustrates some nodes working as a communication hub and carrying maximum data of the whole network. The selection of hub nodes depends on the highest *centrality* or more popularity in the MSNs. Bubble Rap [23] and SimBet [24] are the most prominent examples of social based routing protocols that use hub node as a relay for better transferring of the message. The experimental results define that unfair distribution of load can cause less data packets delivery and huge congestion at hub nodes. Kathiravelu *et al.*[25] and Khabaz *et al.* [26] proved that central node is the best intermediate forwarder because of its central position and ability of receiving more data from other nodes. To explain the congestion problem of a central node the above two works provided a solution for management in storage space. The source node releases forward messages to know about the congestion level of central node before sending the data to the central node.

To distribute the load away from the central node, Radenkovic *et al.* [27] proposed Congestion Aware Forwarding (CAFe). CAFe is based on two main modules; routing and congestion control. The first module is dependent on the social based routing. The algorithm uses *ego betweeness centrality* to transfer data among nodes. The highest *betweeness centrality* defines that the path between the sender and receiver is short. Therefore, the transfer of data takes place without wastage of time. In an ideal scenario, *betweeness centrality* is a better choice for selection of the relay node. However, the drawback of this social property is that it depends on the whole network information, meanwhile such a situation is difficult to achieve in opportunistic networks. Therefore, this scheme uses an *ego* network [28] to calculate an approximate *betweeness centrality*. The second module is used for the congestion control after estimating the statistics of node buffer. The buffer level provides decision either for acceptance or rejection of messages. CAFe calculates the congestion level based on *ego* networks because in opportunistic networks it is difficult to calculate the congestion information of the whole network. The work related to traffic distribution in Soelistijanto *et al.* [29] shows that *ego betweeness centrality* cannot work properly with routing metric in social opportunistic networks. Their results show that *degree centrality* performs better than *ego betweeness centrality* for efficient traffic distribution.

Similar to CAFe, Fair Route [30] provided the solution of unfair load distribution in MSNs. Their forwarding algorithm is based on congestion and routing. In the congestion control module, the queue length of buffer is considered to control congestion. The acceptance or forwarding request depends on the size of the node's queue length. Higher and less queue length of node assigns higher and lower level to nodes, respectively. To estimate a routing decision, the routing module uses the level of interaction with neighbor nodes. The level of interaction can also be represented by *tie strength* that shows how much the number of probabilities for contact in future are. Nevertheless, the drawback of only considering *tie strength* is that, it increases when the node contact increases and decreases exponentially over time. Therefore, for a balanced distribution of traffic, it is necessary to consider node's buffer statistics for forwarding the message.

Adaptive Routing protocol uses degree of connectivity between nodes to calculate the probability metrics as presented by Kathiravelu *et al.* [31]. This scheme defines maximum connectivity between nodes to show that node is popular in the network. Therefore, selection of relay node depends on the nodes' popularities. Nevertheless, the congestion problem in this scheme is protruding due to the maximum number of connections between nodes. To solve this issue Kathiravelu *et al.* proposed Congestion Aware Adaptive (CAA) [25] to cover up the congestion problem. CAA uses advertisement methods to inform the other nodes about its availability of free buffer space. In a first step, every node calculates the degree of connectivity with every neighbor node and then in a second step, it uses popularity level for calculation of safety margin. Furthermore, Wang *et al.* [32] proposed a heuristic algorithm called PRDiscount that used weighted cascade model to work in complex MSN. PRDiscount discovered that degree and other centrality based social properties do not provide the better results to estimate the influential power of node.

Existing methods used centrality concept for selection of relay node and distribution of traffic. For example, Soelistijanto *et al.* [29] proposed that degree centrality provides efficient result for distribution of traffic. Furthermore, Liaqat *et al.* [33] used degree centrality concept to adjust the rate of senders at receiver side and reduced the overhead issue after reduction of acknowledgment packets. Different from the existing works, in this paper, we will use the concept of degree centrality for scheduling the data packet rather than the selection of node or adjustment of data rate. In our scheduling method, the intermediate node prioritizes data packet based on the level of degree centrality.

*2.2. Scheduling in ad-hoc networks*

Scheduling algorithm is helpful to decide which data packet will be served first from the queue to perform efficiently in an ad-hoc environment, when load or congestion of traffic is increasing. Chun *et al.* [34] explains the two types of scheduling algorithms. The first type of scheduling algorithm uses no-priority scheduling (in which FIFO methodology is used). The second type of scheduling algorithm is used for priority settings in data packets. Therefore, multiple scheduling algorithms are available for prioritization in traffic such as weighted-hop scheduling, weighted-distance scheduling, greedy scheduling and round robin scheduling.

The simple priority-scheduling scheme cannot perform efficiently in real networks when the rate of congestion is high, since simple priority algorithms do not have the information of wireless channel capacity from its neighbors. Therefore, Chen *et al.* [35] proposed Congestion Aware Scheduling Scheme (CARE) for MANETs. They used rapid load information that works in a dynamic topology and scarce bandwidth environments. CARE adds one field in both RREQ and RREP to provide the load information. Moreover, it gives high priority to routing packets over data packets and in case of the same priority level, CARE uses FIFO methodology. Sridhar *et al.* [36] presented a new Channel Aware Scheduling for MANETs (CaSMA) algorithm that uses channel's information to take proper action. The working of CaSMA is based on end-to-end channel awareness and it sets channel quality after the estimation of path lifetime. Furthermore, this scheme uses queue size for working in a congested environment, by using schedulable list to provide efficient coordination among end-to-end connections and also provides an ideal schedule to a rough approximation. However, CaSMA has some limitations due to the usage of path lifetime estimation technique. Its performance varies as the accuracy of estimation varies. In some particular scenarios, schedulable list scheme and neighbor management can add overhead with respect to bandwidth consumption.

In terms of work related to considering time varying (fading) channel concepts in MANETs, Ying and Shakkottai. [37] proposed a novel scheme that can work in an environment where global information is difficult to achieve. This algorithm provides advantages in terms of characterization of throughput of the network region with respect to uncertainty of topology and channel. The optimal throughput-scheduling algorithm provides consistency in the network and also provides algorithms that are independent of the network with respect to calculation complexity. Moreover, this scheme also provides delayed information from the local area and mobile instantaneous information. To provide a solution related to congestion control using scheduling method, Akyol *et al.* [38] proposed a scheduling algorithm for MANETs. This algorithm provides a solution to replace the existing Transmission Control Protocol (TCP) in the aspect of today's communication. The performance of the proposed protocol is also compared with TCP. Michele *et al.* [39] explained the capacity region of MANETs and defined the moving process according to a random mobility process. They differentiate the class of scheduling for achieving maximum throughput and multi-commodity flow over an associated contact graph. Their algorithm provides a joint scheduling and routing formulation in addition to enhancement of analysis in heterogeneous nodes with anisotropic mobility.

The existing scheduling algorithms cannot perform efficiently in ASNETs scenario, where the communication of nodes depends on the social properties. In addition, existing methods do not utilize the concept of social properties to set the priority level of data packets for scheduling. The social properties provides advantage in transmission rate and also helpful to solve unfair issue in service rate.

**3. Network model and preliminary**

In order to substantiate congestion issues in ASNETs environment, firstly, we discuss an ASNETs scenario where we consider a network with multiple ($m$) sender and receiver nodes that communicate with a single intermediate node when they are proximal to it. The nodes are divided into two different Social Communities or Groups ($SC_A$ and $SC_B$). In Fig. 1 ASNET communities exploit degree centrality as a social behavior to determine the popularity of nodes. Degree centrality is a social property that determines the popularity of a node in a MSN through its highest number of proximity connections/links. As shown in Fig. 1, $SC_A$ has multiple nodes that are directly or indirectly in relationship within one social community and it also sends data to $SC_B$ nodes. Similar to $SC_A$, the users in $SC_B$ also have direct or indirect relationship within one group but $SC_B$ users work as receivers. $SC_A$ has three sender nodes 1, 2 and 3 that are sending data packets through intermediate node 4. Receiver nodes 5, 6 and 7 are members of $SC_B$ that receive data from node 4. Among the $SC_A$ nodes, node 1 has the highest degree centrality due to the larger number of connections with other node. However, node 2 is defined with an average degree centrality that has less number of connections as compared to node 1. Finally, node 3 has less degree centrality due to less number of connections within $SC_A$. In this scenario, a node with the highest degree centrality can also be called the highest popular one. Moreover, nodes with an average and less degree centrality can also be called an average and less popular nodes, respectively. This proximity based ASNET scenario shows that single intermediate node involves in congestion earlier when $m$ senders transfer data to it. Therefore, a proper scheduling algorithm is required on the intermediate node that should works according to the sender popularity level.

Secondly, we assume that each node knows its own current state of *load* and *social popularity*. Our motivation for selecting load and social popularity based packet scheduling are as follows: 1) the decision to start scheduling the data packets needs *load* consideration. This consideration illustrates how much buffer is already used at intermediate node 4. The reason is that maximum usage of buffer at the intermediate node can convert it into a congested state and cannot provide availability of prioritized or more popular node data packets efficiently. 2) the consideration of social popularity-based assignment to a scheduler shows that most of the nodes are connected or have maximum relationships proximity with this highest popular node. Therefore, we need to avoid the dropping of this highest popular node's data packets. In addition, the earlier transfer of highest popular node's data packets also resolves the congestion issue in an intermediate node.

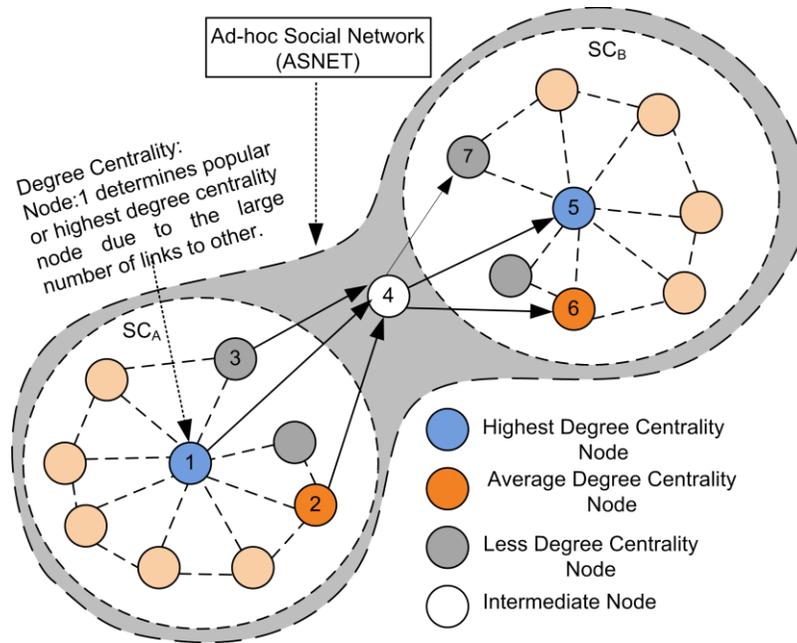

**Fig. 1.** An example demonstrating exploitation of degree centrality social behavior to determine the popularity of nodes.

## 4. User popularity-based packet scheduling

To resolve the congestion issue through scheduling, this section provides the Pop-aware model in detail, as shown in Fig. 2, and scheduling algorithm for ASNETs that considers local congestion information and social popularity (*degree centrality*) information. We begin with describing the importance of considering load and popularity awareness in our algorithm. Moreover, we define the problem and present an approach to its solution. Table. 1 explains the notations with their definitions we use over here.

### 4.1. Load and popularity awareness

In popularity aware scheduling, we should consider two important characteristics: 1) when a node needs to schedule the data packets and 2) which data packet is assigned first to the encountered nodes?

1. To consider the duration of the scheduling process, we study the load of link congestion for various TCP connections to the maximum range of 20 to 50. We find that the level of congestion at the nodes could vary widely. This process demonstrates that if there are *m* flows at any node, the level of the congestion among these connections is unlikely to be similar and can vary over a large range. Therefore, it is useful to consider these congestion levels of the node as a parameter in order to improve performance significantly.

2. The impact of assigning priority to popular data packets is divided into two forms: first, in MSNs, a highest popular sender node has a greater advantage over a less popular node due to the large contacts information. Second, the node transfers high data rate in less delay time and without wastage of bandwidth resources. A node becomes more visible or spectacular than the other nodes, if its tie with the other nodes is high. For example, we consider an ASNETs model in which excessively and less used nodes are available that are communicating each other without any infrastructure. To define the category of resourceful node, an excessively used node has obviously more resources than a less used node in terms of communication capability or more ties with other nodes. In other words, an excessively used node has high *centrality*. To measure the *centrality* three types are available including *degree centrality, closeness centrality* and *betweeness centrality*. In terms of relay for expressing the popularity of an excessively used node, *degree centrality* is broadly used in MSN analysis [40] [41], because it provides connections among individuals and groups.

**Table 1**
Definition of notations/symbols.

| Notation/Symbols | Definition |
|---|---|
| $d^{(pl)}$ | Data packet load |
| $a^{(dr)}$ | Arrival data rate |
| $o^{(dr)}$ | Output data rate |
| $c_a^{(deg)}$ | Degree centrality |
| $deg_a$ | Directly connected neighbors of node $a$ |
| $m$ | Maximum number of connections with sender/intermediate node |
| $\gamma$ | Ratio of service ($0 < \gamma < 1$) |
| $i_a^{(r)}$ | Minimum packet inter-arrival rate |
| $t_a^{(r)}$ | Maximum transmission rate |
| $p_a^{(r)}$ | Level of priority |
| $A$ | $n^{th}$ flow is served at rate $R$ |
| $\alpha(a,R)$ | Remaining amount of workload for flow $a$ that has to be *served* for the queue at any rate $R$. |
| $Sh_a$ | Scheduler of flow |
| $P_{max}$ | Maximum number of data packets |
| $P_{sum}$ | Total number of existing data packets in the queue |
| $P_a$ | Number of existing data packet |

### 4.2. Load state representation

One of the key ideas in Pop-aware is to represent the quality of channel in terms of congestion. The congestion indicator of a node reflects the current channel state. Since the state of channels is changing continuously, nodes have temporal time interval for which they are valid. To provide efficiency in estimation we use the term data packet load ($d^{(pl)}$) to define the load based Active Queue Management (AQM) for congestion detection [42]. The scheduling algorithm should be faster in ASNETs because of dynamic changes in topology and scarceness of bandwidth. The following equation defines $d^{(pl)}$ as a ratio of two parameters: $a^{(dr)}$ and $o^{(dr)}$.

$$d^{(pl)} = a^{(dr)}/o^{(dr)} \tag{1}$$

The $a^{(dr)}$ shows that how much data packets are arrived on node and $o^{(dr)}$ illustrates that how much data packets are transferred from node. To start the scheduling process, we consider queue congestion in terms of load-based information $d^{(pl)}$. The node calculates the $d^{(pl)}$ of the queue. When the node $d^{(pl)}$ reaches to half of the queue, it will start the scheduling process. The next part describes the calculation method of *degree centrality* at each sender node.

### 4.3. Degree centrality representation

To represent the popularity of node, *degree centrality* is broadly used (i.e. defined as the ratio of number of directly connected neighbor's nodes to maximum number of possible communications). For a network that consists on $m$ nodes, the *Degree Centrality* $c_a^{(deg)}$ of a node $a$ is described as:

$$c_a^{(deg)} = deg_a/m-1 = \frac{\sum_{u=1}^{m} x(a,u)}{m-1} \qquad a, u = 1,2,3..... \quad a \neq u \tag{2}$$

In Equation (2), $deg_a$ represents the directly connected neighbors of node $a$. $c_a^{(deg)}$ represents the ratio of connection between the real connected nodes from $a$ to $m$ and divided by $m$-1 that represents the maximum number of connections. To define the most active node in the network, the higher *degree centrality* node possesses larger number of links with other nodes. As such, a central node occupier in the network location may act as a medium for exchange of information between all other nodes in the defined network.

### 4.4. Design problems and notations

The schedule problem is defined as follows. The flow from node $a$ has four parameters $i_a^{(r)}$, $t_a^{(r)}$, $c_a^{(deg)}$, $p_a^{(r)}$ to schedule the data packets. The $i_a^{(r)}$ is the minimum packet inter arrival rate over intermediate link. $t_a^{(r)}$ is the maximum transmission rate of intermediate link. $c_a^{(deg)}$ represents the popularity level of a sender node using (2). $p_a^{(r)}$ represents the level of a node after calculating $d^{(pl)}$, $c_a^{(deg)}$ and service rate (sub-section 4.6). The relation between $c_a^{(deg)}$, $d^{(pl)}$ and $p_a^{(r)}$ is illustrated in Fig. 2. Three senders are sending data to three receivers using single intermediate node. To avoid the congestion on intermediate node, the scheduling algorithm works on it. Fig. 2 demonstrate that every flow is sending data packets to intermediate node buffer with its $c_a^{(deg)}$ value and intermediate node calculates $d^{(pl)}$ on the arrival of every data packet. Furthermore, priority assignment module calculation is based on the $d^{(pl)}$, $c_a^{(deg)}$ and service rate. Let us denote $p_a^{(hr)}$ and $p_a^{(lr)}$ as high and low priority of flow $a$, respectively and sends the data to the output queue according to their priority level. Pair of flows $a$ with social popularity $c_a^{(deg)}$ and queue congestion load $d^{(pl)}$ is defined in the following paragraph. Scheduling Instance $SI$ is defined as a sequence of flows $(a_1, a_2, a_3, \ldots\ldots, a_m)$. The serving criteria of flows are described by the scheduler. Formally, a schedule for $SI$ is described as: $A = (a_1, a_2, a_3, \ldots\ldots, a_m) \cup \{Null\}$.

The function of $A$ shows that $n^{th}$ flow is served at rate $R$ and it is defined as $A(R \subseteq pair(a_n)) = a_n$. Additionally, when the value of $A(R)$ is equal to $Null$, it shows that no flow is served at this time. To explain the criteria of service rate, we have defined it in two ways as $limited\ served$ and $fully\ served$. The node is described as $limited\ served$, if packet of any flow $a$ (except the lowest popular) receives service rate as $A(R_a)=a$. Moreover, when lowest popular node receives this service rate then its packet is called $fully\ served$. In $limited\ served$, the residual service of any flow is defined as $\alpha(a,R)$, remaining amount of workload for flow $a$ that has to be served for the queue at any rate $R$. In the last step, for fair scheduling decision we explain the factors $non$-$served$ and $served$. The $non$-$served$ factor of a schedule is explained as the amount of packets that remains in the networks after assigning the priority to all the flow $\sum_{u=1}^{m} \alpha(u, Rp_u^{(r)})$ and $served\ S(A)$ of schedule is explained as $\sum_{u=1}^{m} fs(u)$. The dependency of $served$ is based on the number of $fully\ served$ packets for each flow. The most common problem for designing a scheduler is to attain maximum $served$ and less $non$-$served$ within particular time period. There are two advantages to reduce the $non$-$served$ packet. Firstly, it resolves the loss related issue that occurs due to congestion and secondly it is helpful to resolve the delay issue that is also due to wastage of resources.

### 4.5. Assignment of packets

For single congested intermediate node with multiple sender flows, we consider a model where a single scheduler is available. When single shared node is utilized by sender flows, the assignment is based on different popularity levels as displayed in Fig. 2. The scheduler $Sh_a$ schedules these flows after considering a single popularity level of $m$ flows that arrive within a specific time interval. Furthermore, the $i_a^{(r)}$ and $t_a^{(r)}$ have same values for all $m$ flows. Accordingly, the maximum number of packets $P_{max}$ are scheduled by $Sh_a$ in highest popularity of any flow $c_h^{(deg)}$. The $c_h^{(deg)}$ defines the maximum popularity level of any sender node that is obtained after the calculation of every sender node $c_a^{(deg)}$.

The $P_{sum}$ represents the total number of existing packets in the queue that is equal to $\sum_{a=1}^{m} P_a$ (where $P_a$ represents the number of existing packets for flow of node $a$). The percentage ratio of throughput is calculated by using $P_{max}/P_{sum}$. To fairly utilize the resources among nodes, we apply ratio technique, $P_a \times (P_{max}/P_{sum})$ which denotes the throughput of each flow of node $a$.

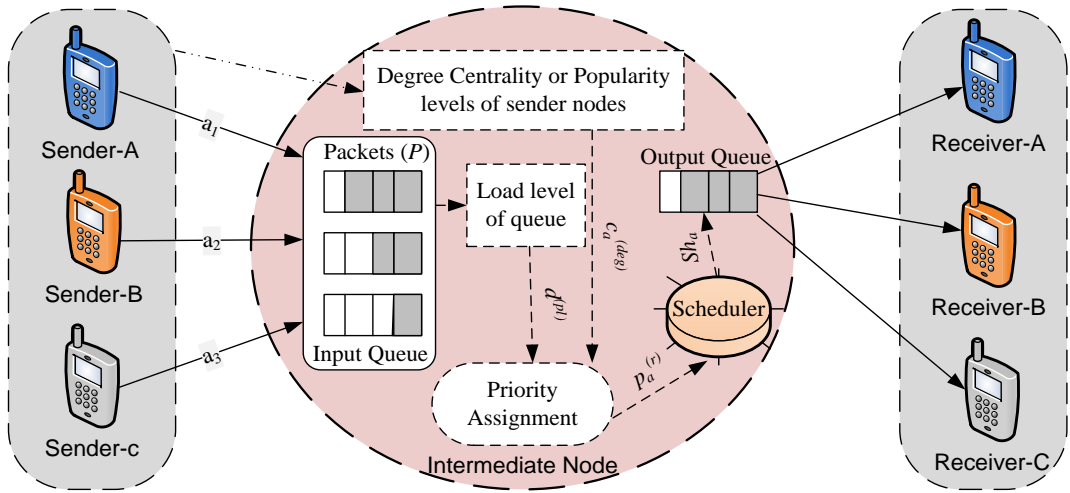

**Fig. 2.** Pop-aware model for socially-aware scheduling.

Additionally, the rate of multiple losses can be describe as $P_a \times (1 - P_{max}/P_{sum})$. The motivation of calculating throughput for scheduler is based on the reality of MSNs, because multiple social applications running within the flows are purely due to the inherit property of ASNETs. Therefore, we need to improve the flow that suffers due to this inherited property of the network. Furthermore, for fair sharing of resources between nodes, the scheduler should be aware of the social properties of nodes.

The next part of this section defines two steps for scheduler. In order to define the scheduling process, first we consider $d^{(pl)}/c_a^{(deg)}$ and secondly, we explain the procedure for scheduled list at the intermediate node.

### 4.6. Priority calculation

To schedule the data packet, our scheme uses priority field that depends on queue *load* of the intermediate node divided by *degree centrality* of the sender node and *active service*. When priority is assigned to any flow, the number of packets served for flow of node $a$ is based on $P_a \times (P_{max}/P_{sum})$. Pop-aware aims to solve two problems: 1) congestion control and 2) efficient utilization of resources. Based on this, we first consider higher priority to flow which has highest social popularity. Second, we consider service received for each flow for the fair utilization of resources. We solve the first problem by using $d^{(pl)}/c_a^{(deg)}$ that is equal to *social rate* in our algorithm and serving queue having lower $d^{(pl)}/c_a^{(deg)}$ values provides higher priority to flows.

For the second problem relying on only $d^{(pl)}/c_a^{(deg)}$ makes it impossible to provide fairness. Therefore, we solve the fairness issue with simple scenario, using $m$ number of flows where each flow for node $a$ has a number of packets $P_a$. Additionally, all data packets service time starts at the same time, the intermediate node has also same $i_a^{(r)}$, $t_a^{(r)}$, and $c_a^{(deg)} = P_a$,( i.e. $P_a$ is based on the degree centrality for each flow). A parameter $\gamma$ (the service ratio of $P_a$) is utilized to explain the ratio of the service within the limited time and the value of $\gamma$ is less than 1 and greater than 0. At this point we highlight that there is no one-to-one mapping between social rate and received service. A lesser social rate for flow of node $a$ as compared to flow of node $u$ does not mean that flow of node $a$ has received less amount of service than flow of node $u$. The variation of $c_a^{(deg)}$s represents the ratio of degree centrality and received service by flow of node $a$. For instance, when two flows have equal $c_a^{(deg)}$s value, then the flow which receives decrease amount of service will demand highest *degree centrality* as compared to the flow having greater amount of service. After using this phenomenon it shows that $d^{(pl)}/c_a^{(deg)}$ distributes the service rate fairly.

The condition where popular flow receives less service needs to be avoided because $c_a^{(deg)}$ varies up to high ratio. This problem can be solved using parameter terms such as active for every flow of node $a$. The *active* parameter for flow of any node $a$ is equal to $P_{max}/P_{sum}$ and it is calculated by using the service rate of packet $a$.

$$\left(P_{max} \times \frac{t_{max}^{(r)}}{i_{max}^{(r)}}\right) \times \left[\frac{P_a \times \frac{t_a^{(r)}}{i_a^{(r)}}}{\sum_{u=1}^{m} P_u \times \frac{t_u^{(r)}}{i_u^{(r)}}}\right] \tag{3}$$

$i^{(r)}_{max}$ and $t^{(r)}_{max}$ show maximum possible $i^{(r)}$ and $t^{(r)}$, respectively. The first part in (3) represents the maximum wireless link rate as well as the maximum data that can be solved. On the other hand, the ratio of flow of node $a$ and the total amount of data that

| **Algorithm 1:** Pseudocode to find lowest $c_l^{(deg)}$ (). |
|---|
| 1: Initialization. |
| 2: Calculation of $c_a^{(deg)}$ for $(a \leftarrow 0,1\ldots\ldots m\text{-}1)$ using Eq. (2); |
| 3: $a \leftarrow 0$,  $deg \leftarrow c_a^{(deg)}$; |
| 4: $l \leftarrow 0$; |
| 5:  Calculation of $c_l^{(deg)}$ () |
| 6:     { |
| 7:         **for** $a < m$ **do** |
| 8:             **if** $c_a^{(deg)} < deg$ |
| 9:                 $deg \leftarrow c_a^{(deg)}$; |
| 10:                 $l \leftarrow a$; |
| 11:             **end if** |
| 12:             $a \leftarrow a+1$; |
| 13:         **end for** |
| 14:     $c_l^{(deg)} \leftarrow 1$; |
| 15:     **return** $c_l^{(deg)}$ |
| 16:     } |

considers all flows of *m* nodes is represented by the second term. The arrival of a new flow or removal of an existing flow can only upgrade the *active* parameter. Both *social rate* and *active service* are considered in order to calculate the priority of flow. The flow is assigned a higher priority in Pop-aware in case *degree centrality* is highest but *social rate* is lesser and *active service* is not yet available.

### 4.7. Pop-aware algorithm

A flow of node *a* is scheduled by the Pop-aware algorithm at the intermediate node. For assigning the prioritization of the data packet, two parameters are needed to be calculated. 1) Using (1), the intermediate node takes a scheduling decision based on its *load* and 2) the *degree centrality* computation using (2) provides highest popular and most prioritized node data packets from connected flows. Three types of nodes are considered in our Pop-aware algorithm: Sender (*S*), Intermediate (*I*) and Receiver (*R*). The above two parameters help building a scheduled list at *I* node and then send it to the *R* node using priority after observing the change in link. In Pop-aware, the $d^{(pl)}$ of *I* node is calculated on the arrival of every new data packet. However, Pop-aware starts scheduling algorithm when the value of $d^{(pl)}$ reaches to the half size of queue otherwise it functions in a FIFO manner. In condition when $d^{(pl)}$ is higher as comparison to half queue length, the intermediate node starts assigning priority to data packets. After assigning priority, Pop-aware selects *k* flows out of *m* flows for scheduling that is helpful for reduction in transmission delay and full utilization of resources. Additionally, when new flow data packet arrives and capacity of queue is full, Pop-aware verifies the defined conditions after finding lowest degree centrality $c_l^{(deg)}$ value of data packet from the intermediate node queue. If new data packet has higher degree centrality value and satisfies the condition then it will be the part of schedulable list otherwise it will drop the arrived data packets. In order to explain the selection methodology of $c_l^{(deg)}$, Algorithm 1 defines the steps of selection of $c_l^{(deg)}$. The selection criteria of *k* flows and new flow data packets are defined in the following paragraphs.

The flow is said to be *fully served*, if none of the packets are queued in the intermediate node. To solve the *fully served* scheduling algorithm, the following two problems are considered. The first problem provides a solution to calculate the number of data packets that are schedulable. The second problem aims to calculate the priority for the new flow of node *n* and decide whether it can be a part of the existing schedulable set. The details are shown in Algorithm 2.

To solve the above two problems: first, the schedulable set that chooses *k* flows out of *m* number of flows at intermediate node is considered. Pop-aware resolves this issue after utilizing the real-time scheduling standard result [43]. The minimum inter-arrival

---

**Algorithm 2** Pseudocode of Pop-aware for scheduling.

1: At the beginning of each frame
2: **for all** arrival data packet **do**
3:     Calculate the $d^{(pl)}$ of *I* node using Eq. (1);
4:     **if** ($d^{(pl)}$>*queue size*/2)
5:         Calculate the *Degree Centrality* $c_a^{(deg)}$ of each connected *S* node using Eq. (2);
6:         // **Assigning priority to flow for scheduling**
7:         **for all** new flow arrival or existing flow leave **do**
8:             Set priority using $d^{(pl)}/c_a^{(deg)}$ and suitable *service* using Eq. (3);
9:         **end for**
10:         // **Considering *k* flow out of *m* flow for initially transferring**
11:         **for** (*a* =1 to *a* ≤ *k* ) **do**
12:             Select *k* flow out of *m* flow after satisfying increment to decrement priority and Eq. (4);
13:         **end for**
14:         **for** arrival of new flow *n* data packet and the capacity of queue is filled **do**
15:             Calculate the $c_l^{(deg)}$ value from the intermediate node queue using Algorithm 1;
16:             **if**($c_n^{(deg)}$>$c_l^{(deg)}$) **and** ($t_n^{(r)}/i_n^{(r)}$)≤($t^{(r)}/i^{(r)}$)**then**
17:                 *n* join the scheduling list after satisfying Eq. (4);
18:             **else**
19:                 Drop data packet;
20:             **end if**
21:         **end for**
22:     **else**
23:         Apply FIFO method to queue
24:     **end if**
25: **end for**

rate of packet from flow $a$ is denoted by $i_a{}^{(r)}$, which required $t_a{}^{(r)}$ units of rate for transmission. The rate of all flow from node $a$ should be less than or equal to 1 as depicted in (4).

$$\sum_{a=1}^{k} \left\{ \frac{t_a{}^{(r)}}{i_a{}^{(r)}} \right\} \leq 1 \tag{4}$$

Equation. (4) can be redesigned according to the $c_a{}^{(deg)}$ based rate adjustment. In which $c_a{}^{(deg)}/ i_a{}^{(r)}$ denotes the arrival rate of data packets from flow $a$ that required $c_a{}^{(deg)}$ $(t_a{}^{(r)}/i_a{}^{(r)})$ rate for transmission. However, the rate of adjustment of all flows should be less than or equal to $P_{max}$. Generally, the selection of $k$ flows out of $m$ number of flows can be done through many possible ways. One of them is based on the values of $i_a{}^{(r)}$ and $t_a{}^{(r)}$. However, this dependency gets a smaller value of $k$ at smaller value of $i_a{}^{(r)}$. On the other hand, *degree centrality* methodology is used to choose the value of $k$. *Degree centrality* based scheduling provides high advantage in ASNETs scenarios, where data flow is dependent on social properties (Popularity). Therefore, based on the result of *degree centrality*, the highest popular node data packet is selected first and served. Moreover, it arranges all the $m$ number of flows in terms of the decreasing priority value. The first $k$ flows which are related to the schedulable set are selected using this arrangement. The intermediate node uses (4), which depends on the rates, to select the $k$ flows for scheduling.

To solve the second issue, $Sh$ schedulable at a particular node is defined. Let us assume the lowest social popularity in the schedulable set denoted by $c_l{}^{(deg)}$ of flow for node $l$. Here $l$ denotes the flow of node that has lowest priority in the existing schedulable queue. Therefore, to add new data packet in the schedulable list it is necessary to compare the new data with already existing data. The $c_n{}^{(deg)}$ is schedulable if $c_n{}^{(deg)}$ of new flow $n$ is greater than $c_l{}^{(deg)}$ of available flow $l$ and the $(t_n{}^{(r)}/i_n{}^{(r)})$ for $n$ flow is lesser than or equal to $(t_l{}^{(r)}/i_l{}^{(r)})$ of available flow $l$. The reason of smaller value of $(t_n{}^{(r)}/i_n{}^{(r)})$ as compared to $(t_l{}^{(r)}/i_l{}^{(r)})$ shows that it does not violate the condition which is defined in (4) for selection of $k$ flows. In case, if the above objective is not achieved, it drops the arrival flow data packet.

## 5. Performance analysis

To define the theoretical analysis of Pop-aware scheme, here we first discuss the setting of problem with the description of suppositions. Subsequently, we will define the performance of our scheme using this setting. This analysis utilizes the concept of popularity level information of each sender node that is involved in the defined ASNETs.

Let us suppose there are $m$ numbers of total nodes in ASNETs. In this scenario, multiple senders are sending data to a single intermediate node. According to this scenario, intermediate node involves in congestion at an earlier stage. This is because, the nodes in ASNETs have limited number of queue capacity and the range of receiving data packet is higher than its queue capacity. In order to resolve this issue, a proper mechanism is required to schedule the data packet according to the concept of socially-aware networks. Therefore, in this analysis we tried to resolve the above mentioned issue using problem definition and some assumptions. The maximum number of data packets that can be stored in queue capacity of an intermediate node is denoted by $P_{max}$ and each node has its own *degree centrality* $c_a{}^{(deg)}$ value. The value of $c_a{}^{(deg)}$ indicates the popularity level information of node that is also helpful to analyze the minimum inter-arrival rate $i_a{}^{(r)}$ of data packets. Furthermore, $c_a{}^{(deg)}$ is also helpful in sustaining or dropping the data packets. In our systematic analysis, load of data packet $d^{(pl)}$ is also an important parameter. This is because, in ASNETs scenario, node queue congestion is the main reason for dropping of data packets. Therefore, exact consideration of load is necessary to start the scheduling process and avoiding the dropping of data packets of highest popular node. To resolve the congestion problem in an intermediate node, firstly, intermediate node should know how to schedule all data packets in a queue. Secondly, intermediate node should also know which data packet should drop first after the arrival of new data packet, when the capacity of its queue is full.

The considerations of above two issues are helpful to increase the transmission range and reduction in delay. Therefore, in next sub-sections we explained the mathematical solution of Pop-aware after considering of vital metrics such as average transmission rate with average delay. Pop-aware utilizes the concept of *degree centrality* and derives this social property value for each data packet. According to the *degree centrality* based social property, our scheme performs two main functions. First, to perform the scheduling function, our scheme arranges all data packets from higher to lower order that is based on source popularity level. Second, in the queue management function, Pop-aware takes a decision for dropping of data packet. This function will be performed when the new data packet is arrived in an intermediate node and the load capacity of it's queue is full. In this case, intermediate node will drop the data packet that has lowest popularity.

### 5.1. Increment in average transmission rate

In this sub-section, we checked the performance of average transmission rate using the environment where each data packet has minimum $i_a{}^{(r)}$ value based on its node's social popularity level. In order to increase the average transmission rate of data packets in an ASNET, our scheme will utilize the concept of social property ($c_a{}^{(deg)}$) of node for data packets. Therefore, in our theorem 1 we derived the performance of ASNET using proper queue management and scheduling.

***Theorem 1:*** To quantify the problem here, we suppose that there are $P_{sum}$ data packets in an intermediate node which has maximum transmission rate $t_a^{(r)}$ of the flow $a$. However, the $t_a^{(r)}$ of each flow depends on the $c_a^{(deg)}$ level and the range of each data packet $a \in [1, P_{sum}]$. The increment in transmission rate is based on the selection criteria of flows. Therefore, here we consider two suppositions. 1) Let us suppose that the $k_a (t_a^{(r)}/i_a^{(r)})(c_a^{(deg)})$ be the maximum number of data packets from the selected flows for scheduling. The criteria of selection of $k$ flows are based on the *degree centrality* level. However, the total range of data packets of $k$ flows should be less than equal to $P_{max}$. 2) In addition, we suppose that the arrival of new flow is denoted by $n_a (t_a^{(r)}/ i_a^{(r)})( c_a^{(deg)})$ that is included in schedulable list after comparison with selected $k$ flows. To increase the transmission rate, following equation is applied to each data packet.

$$S_a^{tr} = \left[ \left(1 - \frac{n_a\left(\frac{t_a^{(r)}}{i_a^{(r)}}\right)c_a^{(deg)}}{m-1}\right) d^{(pl)} \alpha(a, R) \exp\left(-d^{(pl)} k_a \left(\frac{t_a^{(r)}}{i_a^{(r)}}\right) c_a^{(deg)} \; \alpha(a, R)\right) \right] \tag{5}$$

*Proof*: In ASNETs, the load of data packets in an intermediate node is distributed exponentially with variable $d^{(pl)}$. The possibility is, that data packet of selected $k_a (t_a^{(r)}/ i_a^{(r)})(c_a^{(deg)})$ flow will not be transferred from an intermediate node. It is illustrated that the probability of next arrival load of node becomes higher than the remaining amount of workload for selected flow $a$, represented as $\alpha(a,R)$. The defined statement is equal to exp(-$d^{(pl)}\alpha(a,R)$). We know that data packet $a$ is selected for transmission by using $k_a (t_a^{(r)}/ i_a^{(r)})(c_a^{(deg)})$. However, here we are considering that the selected data packet has not been transferred yet. According to that, we can derive the equation of probability that the data packet itself will not be transferred.

$$P(\text{packet } a \text{ not transferred} \mid \text{not transferred yet}) = \prod_{a=1}^{k_a\left(\frac{t_a^{(r)}}{i_a^{(r)}}\right)c_a^{(deg)}} \exp\left(-d^{(pl)}\alpha(a, R)\right)$$

$$= \exp\left(-d^{(pl)} k_a \left(\frac{t_a^{(r)}}{i_a^{(r)}}\right) c_a^{(deg)} \; \alpha(a, R)\right) \tag{6}$$

There is no consideration about selection of given data packet $a$, that may be created in the future through new arrival of node. There is also no consideration that a selected data packet $a$ could be dropped within $\alpha(a,R)$. The problem becomes difficult due to supposition of future upcoming events. However, the similar suppositions are applicable for all data packets uniformly and therefore can give explanation of association between the transmission probabilities for different data packets.

Further, we are also required to consider that what has happened in the intermediate node queue, since the data packet has been already transferred. It is provided that all source nodes have equal opportunity to send the data packets, the probability that a data packet $a$ has been already transferred and consideration of new data packet, is defined in following equation.

$$P(\text{packet } a \text{ already transferred}) = \frac{n_a\left(\frac{t_a^{(r)}}{i_a^{(r)}}\right)c_a^{(deg)}}{m-1} \tag{7}$$

The sum of (6) and (7) illustrates the probability of data packet $a$ getting transferred before the arrival of $i_n^{(r)}$:

$$P_a = P(\text{packet } a \text{ not transferred yet}) \times \left(1 - \exp\left(-d^{(pl)} k_a \left(\frac{t_a^{(r)}}{i_a^{(r)}}\right) c_a^{(deg)} \; \alpha(a, R)\right)\right) +$$

$$P(\text{data packet } a \text{ already transferred})$$

$$P_a = \left(1 - \frac{n_a\left(\frac{t_a^{(r)}}{i_a^{(r)}}\right)c_a^{(deg)}}{m-1}\right) \times \left(1 - \exp\left(-d^{(pl)} k_a \left(\frac{t_a^{(r)}}{i_a^{(r)}}\right) c_a^{(deg)} \; \alpha(a, R)\right)\right) + \frac{n_a\left(\frac{t_a^{(r)}}{i_a^{(r)}}\right)c_a^{(deg)}}{m-1}$$

The total transmission rate for the overall network is defined in the following equation:

$$T^{tr} = \sum_{a=1}^{P_{sum}} \left[\left(1 - n_a \left(\frac{t_a^{(r)}}{i_a^{(r)}}\right) c_a^{(deg)}/m-1\right) \times \left(1 - \exp\left(-d^{(pl)} k_a \left(\frac{t_a^{(r)}}{i_a^{(r)}}\right) c_a^{(deg)} \; \alpha(a, R)\right)\right) + n_a \left(\frac{t_a^{(r)}}{i_a^{(r)}}\right) c_a^{(deg)}/m - 1\right]$$

When transfer of data packet is limited and the queue of intermediate node is congested, ASNET node should take decision to drop a lowest popular node data packet which also leads to the best achievement in the overall transmission rate $T^{tr}$. For best achievement in the decision, we differentiate $T^{tr}$ with respect to $k_a (t_a^{(r)}/ i_a^{(r)})(c_a^{(deg)})$.

$$\Delta(T^{tr}) = \sum_{a=1}^{P_{sum}} \left[\left(\partial P_a/\partial \left[k_a \left(\frac{t_a^{(r)}}{i_a^{(r)}}\right) c_a^{(deg)}\right]\right) \times \Delta k_a \left(\frac{t_a^{(r)}}{i_a^{(r)}}\right) c_a^{(deg)}\right]$$

$$\Delta(T^{tr}) = \sum_{a=1}^{P_{sum}} \left[ \left(1 - n_a \left(\frac{t_a^{(r)}}{i_a^{(r)}}\right) c_a^{(\text{deg})}/m - 1\right) d^{(pl)} \alpha(a,R) \exp\left(-d^{(pl)} k_a \left(\frac{t_a^{(r)}}{i_a^{(r)}}\right) c_a^{(\text{deg})} \alpha(a,R)\right) \times \right.$$
$$\left. \Delta k_a \left(\frac{t_a^{(r)}}{i_a^{(r)}}\right) c_a^{(\text{deg})} \right]$$

The aim of our Pop-aware scheme is to increase ($T^{tr}$). However, in this situation when data packet drops, for example, we know that $\Delta k_a (t_a^{(r)}/i_a^{(r)})(c_a^{(deg)})$ will be reduced if we drop an existing data packet from the intermediate node queue, $\Delta k_a (t_a^{(r)}/i_a^{(r)})(c_a^{(deg)})$ will be constant if we won't drop an existing data packet from the intermediate node queue, and $\Delta k_a (t_a^{(r)}/i_a^{(r)})(c_a^{(deg)})$ will be increased if we save the newly arrived data packet. On the basis of this, Pop-aware assigns priority to data packets using each data packet's social popularity in (5). It is also helpful to schedule and drop the data packets. The selection of data packet $a$ based on highest degree centrality value, increases the overall transmission rate.

### 5.2. Decrement in average delay

Our further consideration is to minimize the average delay in transmission. Here, now we suppose that all data packets have transmission rate based on its *degree centrality*. The delay of intermediate node will be reduced after transferring high loaded data packet first and using proper scheduling decision method. In order to obtain the best achievement in data packets with less delay time following theorem explains the steps. The setting of theorem 2 is same as 1.

**Theorem 2:** For decrement in transmission related delay of all data packets, ASNET node should apply the popularity aware policy using the following function for each data packet $a$.

$$S_a^{dt} = 1 \Bigg/ d^{(pl)} k_a \left(\left(\frac{t_a^{(r)}}{i_a^{(r)}}\right) c_a^{(\text{deg})}\right)^2 \left(1 - \frac{n_a \left(\frac{t_a^{(r)}}{i_a^{(r)}}\right) c_a^{(\text{deg})}}{m-1}\right) \tag{8}$$

*Proof:* We have represented the delay in transmission for data packet $a$ as random value $RV_a$. In a case, if the data packet has already been transmitted, then the value of delay is equal to zero. The total predictable delay in transmission $T^{dt}$ for which selection of flow in the intermediate node of all data packets is given in (9).

$$T^{dt} = \sum_{a=1}^{P_{sum}} \left[ \left(n_a \left(\frac{t_a^{(r)}}{i_a^{(r)}}\right) c_a^{(\text{deg})}/m - 1\right) \times 0 + \left(1 - \left(n_a \left(\frac{t_a^{(r)}}{i_a^{(r)}}\right) c_a^{(\text{deg})}/m - 1\right)\right) \times E\left(RV_a | RV_a > \frac{t_a^{(r)}}{i_a^{(r)}} c_a^{(\text{deg})}\right) \right] \tag{9}$$

The first selection of data packets $a$ transferred from an intermediate node follows an exponential distribution with mean $1/k_a (t_a^{(r)}/i_a^{(r)})(c_a^{(deg)}) d^{(pl)}$. It is given as

$$E\left(RV_a | RV_a > \frac{t_a^{(r)}}{i_a^{(r)}} c_a^{(\text{deg})}\right) = \left(\frac{t_a^{(r)}}{i_a^{(r)}}\right) c_a^{(\text{deg})} + 1 \Bigg/ d^{(pl)} k_a \left(\frac{t_a^{(r)}}{i_a^{(r)}}\right) c_a^{(\text{deg})} \tag{10}$$

After replacement of (10) in (9), we have result in (11).

$$T^{dt} = \sum_{a=1}^{P_{sum}} \left[ \left(1 - \left(n_a \left(\frac{t_a^{(r)}}{i_a^{(r)}}\right) c_a^{(\text{deg})}/m - 1\right)\right) \left(\frac{t_a^{(r)}}{i_a^{(r)}}\right) c_a^{(\text{deg})} + 1/d^{(pl)} k_a \left(\frac{t_a^{(r)}}{i_a^{(r)}}\right) c_a^{(\text{deg})} \right] \tag{11}$$

In next step, differentiate $T^{dt}$ with respect to $k_a (t_a^{(r)}/i_a^{(r)})(c_a^{(deg)})$ that is helpful in searching the scheme that increase in $T^{dt}$.

$$\Delta(T^{dt}) = \sum_{a=1}^{P_{sum}} \left[ \left\{ 1/d^{(pl)} k_a \left(\left(\frac{t_a^{(r)}}{i_a^{(r)}}\right) c_a^{(\text{deg})}\right)^2 \right\} \left\{ \left(n_a \left(\frac{t_a^{(r)}}{i_a^{(r)}}\right) c_a^{(\text{deg})}/m - 1\right) - 1 \right\} \times \Delta k_a \left(\frac{t_a^{(r)}}{i_a^{(r)}}\right) c_a^{(\text{deg})} \right] \tag{12}$$

According, to the above equation, the efficient dropping or transmission selection will be the one that increases (-($T^{dt}$)). This provides results to each data packet social popularity of Eq. (8).

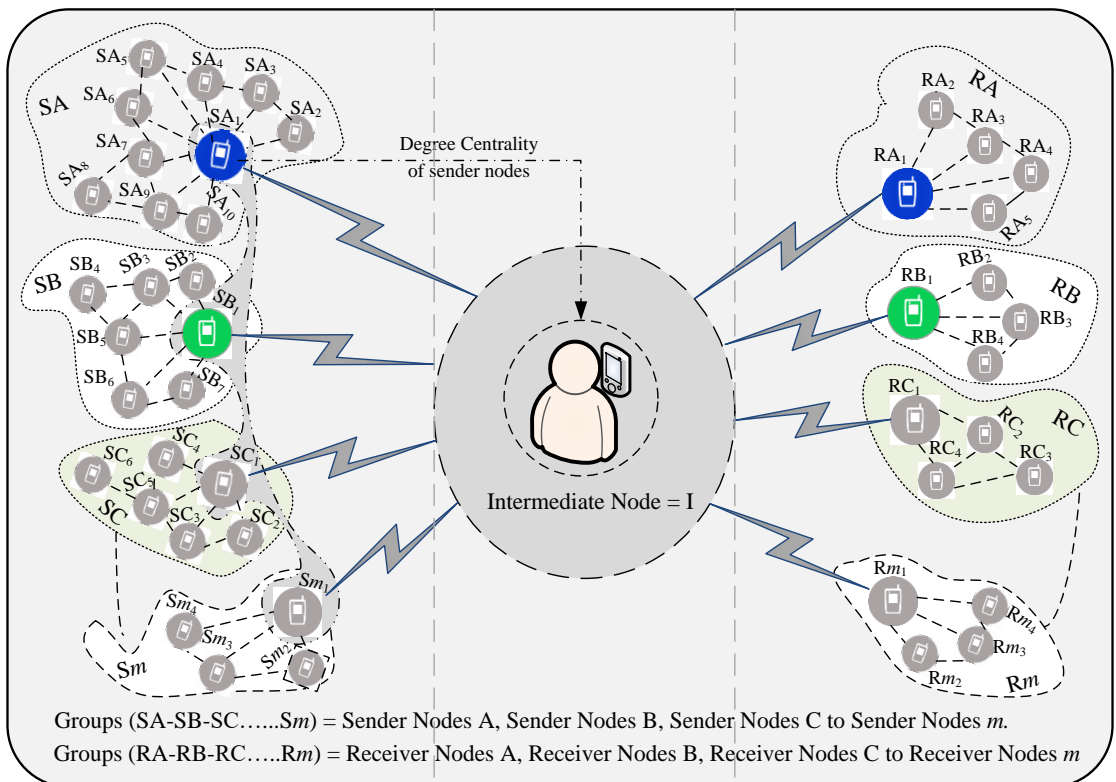

**Fig. 3.** Simulation setup for ASNETs' users with degree centrality based communication.

## 6. Evaluation and discussion

The first part of this section provides the simulation setup and methodology and the second part gives an understanding of *degree centrality* concept using the calculation methodology as shown in Fig. 3. In the third part of this section, we evaluate the performance of our proposed scheduling scheme Pop-aware by comparing it with the existing algorithms.

### 6.1. Simulation setup and methodology

The main objective of Pop-aware algorithm is to implement a scheduling-based algorithm for differentiating data packets of real-time load, while sustaining efficient resource distribution in accordance to degree centrality level of users. The proposed algorithm is evaluated through the OPNET simulation tool that provides substantiation of Pop-aware algorithm quality when degree centrality social property is used for communication. The performance of Pop-aware is verified after avoiding the dropping of data packets of popular node. In addition, to reduce the congestion level on intermediate node with high delivery rate, our algorithm schedules packets based on source node popularity level. The scheduling of packets based on popularity is also helpful to reduce the transmission delay and loss rate.

The comparison of Pop-aware is done with other scheduling algorithms that are called CaSMA [36], (S-R-1) proposed by Ying and Shakkottai [37], and wGPD [38]. The motive of Pop-aware evaluation is to validate whether it works efficiently as compared to other scheduling algorithms in terms of control overhead, total overhead, average throughput, packet loss rate, average delay and packet delivery rate. We consider a degree centrality based scheduling algorithm for efficient utilization of resources and reducing the congestion on an intermediate node. The consideration of degree centrality-based assignment to a scheduler shows that most of the nodes are connected or maximum relationships are proximal to the highest popular node. Therefore, we need to avoid the dropping of this highest popular node's data packets. Furthermore in simulation setup, senders contain degree centrality level that indicates popularity based communication and intermediate node schedule packets based on highest degree centrality level.

Fig. 3 defines the ASNET scenario that is used for simulation setup. To perform in an congested environment and verify the results, we set single intermediate node with multiple source nodes in our simulation setup. Moreover, it includes 802.11 MAC protocol in which the capacity of wireless channel is 2 Mb/sec. In our scenario, we considered 50 nodes that work as a source nodes and individual node have 250 m radio transmission range in the area of 1000 m × 1000 m that placed randomly. The simulation runs multiple time and the collected values results are based on the average data. The network includes object such as ad-hoc gateway, ad-hoc nodes, and definition of application object. Static routing is selected as a routing protocol for our simulation scenario. We set a scheduler buffer size of 64 packets for each node. The time for running each simulation is 800 seconds. The data source for each node is set as Constant Bit Rate (CBR), the transmission rate for each source node is a certain rate and the size of packet is 512 bytes. In the simulation environment, the load of traffic is variable and its dependency is based on the number of connections or the sending rates of packets. The setting of nodes in one social group and the calculation methodology of popularity of nodes are defined in the next sub-section.

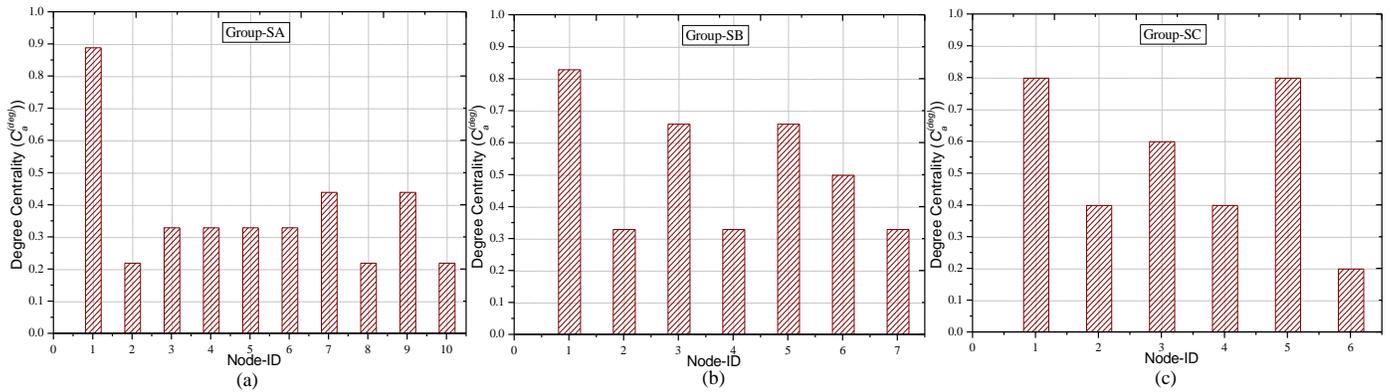

**Fig. 4.** Impact of degree centrality $c_a{}^{(deg)}$ on nodes connection with in group (a) SA, (b) SB and (c) SC.

### 6.2. Calculation of $c_a{}^{(deg)}$ based on pop-aware scenario

To design an ASNETs scenario, we explain our own social graph as shown in Fig. 3 that represents social communication between sender and receiver nodes. Fig. 1 also explains why *degree centrality* is useful in our scenario. In our scenario, the network consists of approximately fifty nodes with every node having its own group. The node is called popular node, if it has the maximum relationship with the other nodes in ASNETs. The consideration of *degree centrality* provides two advantages: 1) it provides full utilization of resources in terms of bandwidth because this social property shows that node carry maximum data of the other node within a group due to high relationship with them and, 2) it gives high advantage to reduce the congestion at intermediate node due to its proper scheduling scheme and transfers most popular data packets first.

Fig. 3 shows that nodes are divided in different groups (A, B, C… *m*). The figure explains that $SA_1$, $SB_1$, $SC_1$…. $Sm_1$ are sender nodes that are communicating with *I* node. The selections of sender nodes from groups are based on the coverage area between sender nodes and an intermediate node. The congestion can occur at the intermediate node in the given scenario. In response, the intermediate node may drop popular node data packets due to the large number of sender nodes and improper scheduling decision. Therefore, we select *degree centrality* concept to transfer the popular node data packets first for scheduling. Table 2 explains the calculation of node $c_a{}^{(deg)}$ (using (2)) and $c'_a{}^{(deg)}$ from Fig. 3. $c_a{}^{(deg)}$ and $c'_a{}^{(deg)}$ are used to calculate the popularity level of sender groups from *A* to *m*. $c'_a{}^{(deg)}$ only shows a partial function of the size of the network where it is calculated. The calculation of $c'_a{}^{(deg)}$ is as follows:

$$c'^{(deg)}_a = \sum_{u=1}^{m} x(a,u) \tag{13}$$

**Table 2**
Calculations of degree centrality for *m* nodes in A, B and C groups.

| Node-ID | Group-SA | | Group-SB | | Group-SC | |
|---|---|---|---|---|---|---|
| | $c'_a{}^{(deg)}$ | $c_a{}^{(deg)}$ | $c'_a{}^{(deg)}$ | $c_a{}^{(deg)}$ | $c'_a{}^{(deg)}$ | $c_a{}^{(deg)}$ |
| 1 | 8 | 0.89 | 5 | 0.83 | 4 | 0.8 |
| 2 | 2 | 0.22 | 2 | 0.33 | 2 | 0.4 |
| 3 | 3 | 0.33 | 4 | 0.66 | 3 | 0.6 |
| 4 | 3 | 0.33 | 2 | 0.33 | 2 | 0.4 |
| 5 | 3 | 0.33 | 4 | 0.66 | 4 | 0.8 |
| 6 | 3 | 0.33 | 3 | 0.5 | 1 | 0.2 |
| 7 | 4 | 0.44 | 2 | 0.33 | - | - |
| 8 | 2 | 0.22 | - | - | - | - |
| 9 | 4 | 0.44 | - | - | - | - |
| 10 | 2 | 0.22 | - | - | - | - |

However, in some application scenarios, this is inappropriate. Therefore, these applications use the absolute degree of calculation i.e. $c_a^{(deg)}$ using (2). In addition, group wise *degree centrality* of every node is shown in Figs. 4(a)-4(c). To explain the concept of social popularity of node, we consider three groups. Group-SA$_1$ shows that it has the highest *degree centrality* or popular node than Group-SB$_1$ and SC$_1$. The calculated result illustrates that SA$_1$ has the highest priority; thus helpful in solving the congestion issue. For further understanding, we are also explaining the calculation methodology of SA, SB and SC nodes. In SA group total numbers of nodes are 10 and SA$_1$ has 8, SA$_2$ has 2, and SA$_3$ has 3 direct connections with other nodes. Furthermore, in SB group total numbers of nodes are 7 in which SB$_1$ has 5, SB$_2$ has 2, and SB$_3$ has 4 direct connections with other nodes. Finally, in SC group total numbers of nodes are 6 in which SC$_1$ has 4, SC$_2$ has 2, and SC$_3$ has 3 direct connections with other nodes.

$c_{SA_1}^{(deg)} = 8/10\text{-}1 = 0.89$

$c_{SA_2}^{(deg)} = 2/10\text{-}1 = 0.22$

$c_{SA_3}^{(deg)} = 3/10\text{-}1 = 0.33$

$c_{SB_1}^{(deg)} = 5/7\text{-}1 = 0.83$

$c_{SB_2}^{(deg)} = 2/7\text{-}1 = 0.33$

$c_{SB_3}^{(deg)} = 4/7\text{-}1 = 0.66$

$c_{SC_1}^{(deg)} = 4/6\text{-}1 = 0.80$

$c_{SC_2}^{(deg)} = 2/6\text{-}1 = 0.40$

$c_{SC_3}^{(deg)} = 3/6\text{-}1 = 0.60$

### 6.3. Simulation results and analysis

As depicted in Figs. 5(a) and (b), we first compare our approach (Pop-aware) with the other scheduling schemes, wGPD, S-R-1 and CaSMA based on control overhead and the total overhead. Control overhead is the total number of control packets transmitted by the sender for delivering the data packets to the receivers. Total overhead is the total number of data and control packets transmitted for delivering the total number of data packets. Figs. 5(a) and (b) illustrate the overhead of packets from minimum to the maximum number of connections (1-50) and the standard deviation of average values are also shown in figures. From the

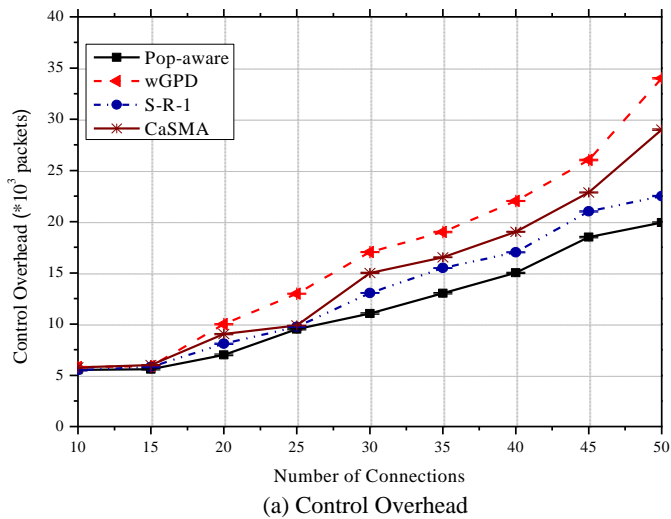

(a) Control Overhead

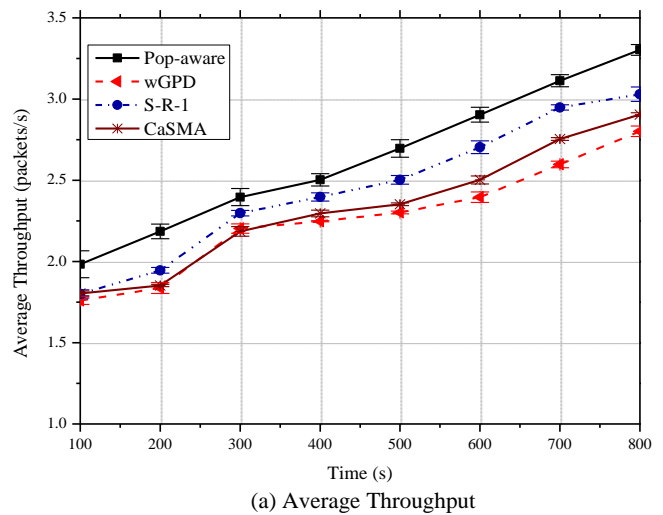

(a) Average Throughput

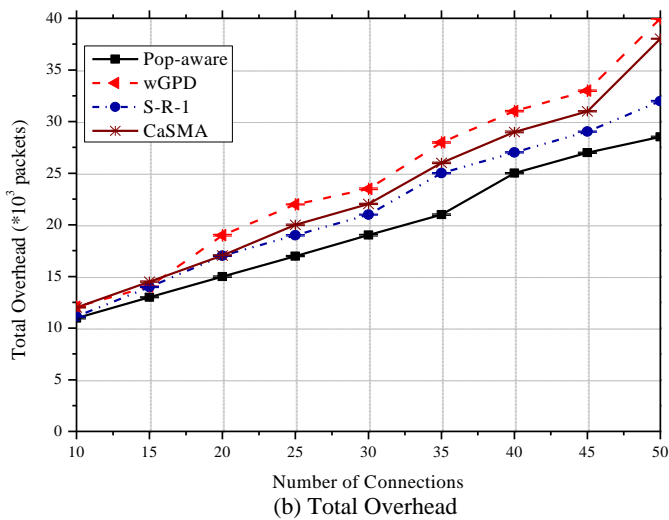

(b) Total Overhead

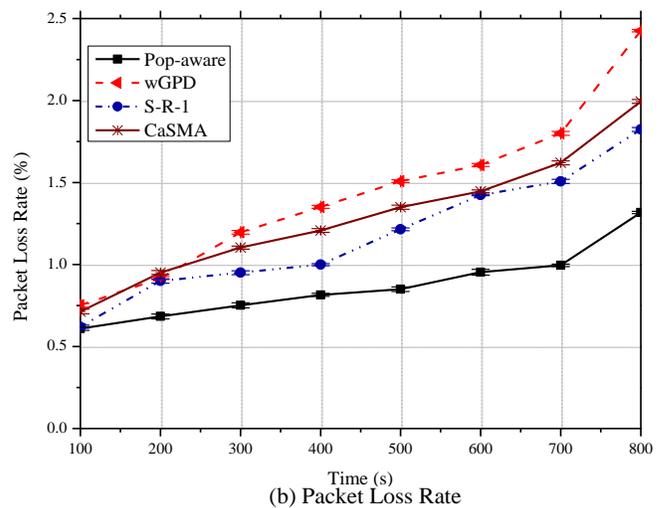

(b) Packet Loss Rate

**Fig. 5**. Performance with respect to number of connections

**Fig. 6**. Performance with respect to simulation time

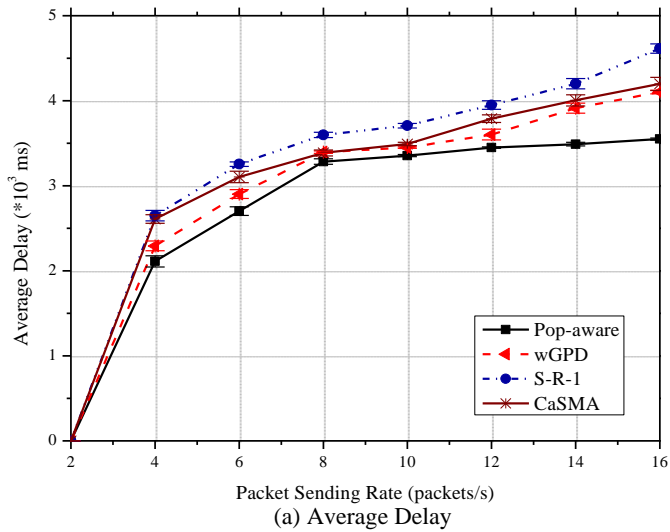

(a) Average Delay

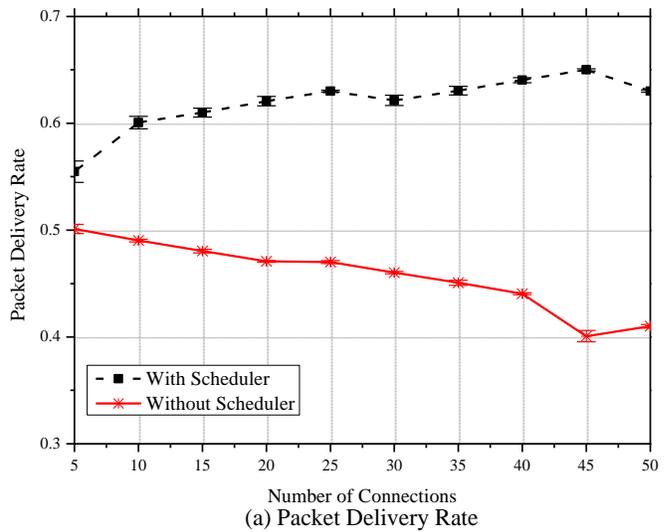

(a) Packet Delivery Rate

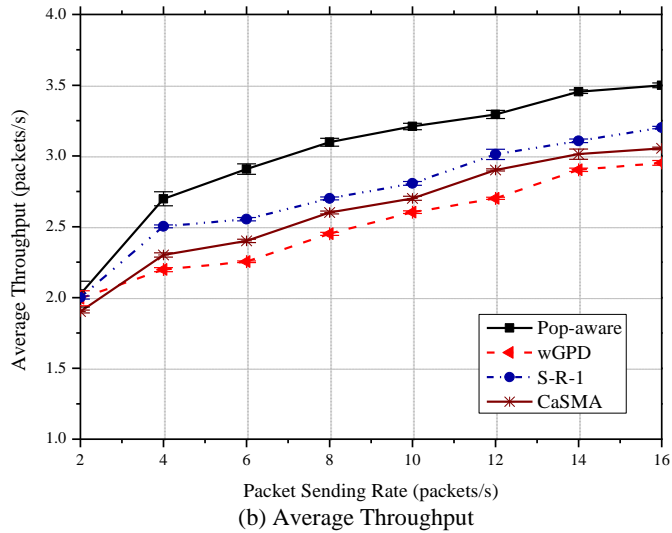

(b) Average Throughput

**Fig. 7.** Performance with respect to packet sending rate

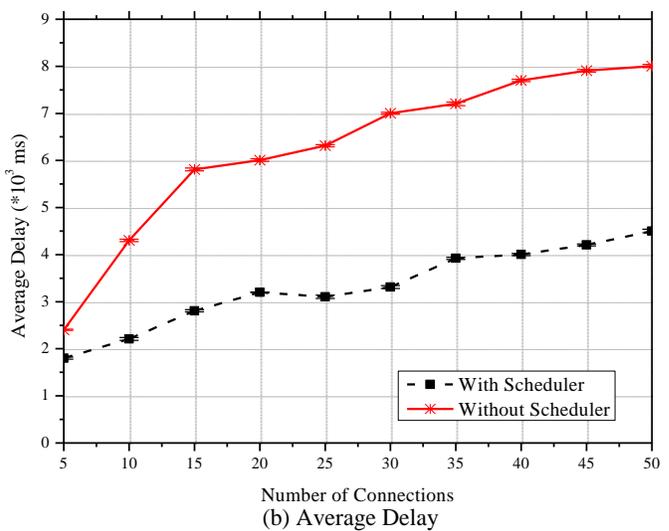

(b) Average Delay

**Fig. 8.** Significance of scheduling with respect to number of connections

figures, wGPD scheme performs worst in case of control and total overhead. In all overheads, there is several numbers of connections with high amount of load, while there is also several numbers of connections with low amount of load. The overhead of all the methods start growing after $10^{th}$ number of connections. After that, the maximum range of control overhead of Pop-aware is $19*10^3$, S-R-1 is $22*10^3$, CaSMA is $29*10^3$ and wGPD is $34*10^3$ as displayed in Fig. 5(a). Moreover, the maximum range of total overhead as depicted in Fig. 5(b) of Pop-aware is also less as compared to existing schemes. These results show that our proposed scheme incurs less overhead due to its reactive approach. The reactive approach works when the size of the queue is near to congestion. The early start of scheduling process creates high overhead in network. However, late start of scheduling process involves the network in congestion state and the scheduling time of data packet is also not enough. Therefore, we use half size of queue to start the scheduling process.

Figs. 6(a) and (b) illustrate the performance comparisons of the throughput and the loss rate of packets with respect to time and the standard deviation of average values are also shown in figures. Fig. 6(a) shows that as the duration of time increases, it incurs higher throughput for all the scheduling schemes (Pop-aware, wGPD, S-R-1 and CaSMA). The logic behind this is that when more nodes are connected, the average throughput of the nodes reaches the maximum position. The Pop-aware scheduling algorithm depicts higher average throughput in comparison to other scheduling algorithms. The reason for higher average throughput of Pop-aware scheme is due to the initial transfer of most popular node data packets. However, the difference of average throughput between wGPD, S-R-1 and CaSMA is not too much. It shows almost the same average throughput from 100 to 300 s, while our proposed method achieves 9.1%, 12.1% and 16.67% compared to S-R-1, CaSMA and wGPD, respectively. Fig. 6(b) illustrates that Pop-aware reduces packet loss rate after behaving separately with different packet flows. Existing schemes use dropping mechanisms to reduce the congestion, due to this, loss rate is higher. However, here we use the proper scheduling methodology in response of dropping to reduce the congestion in the network. The percentage of loss rate of Pop-aware is less as compared to wGPD, S-R-1 and CaSMA. Pop-aware scheduling scheme uses degree centrality and load based prioritization for flow. Therefore, the percentage of our scheme loss rate is 0.5%, 0.7%, and 1.1% less in comparison to S-R-1, CaSMA and wGPD, respectively.

Figs. 7(a) and (b) illustrate the effect of packet sending rate on the average delay and average throughput and the standard deviation of average values are also illustrated in figures. As depicted in Fig. 7(a), the average delay of Pop-aware, wGPD, S-R-1 and CaSMA scheduling schemes are slightly different when the load of traffic is high. The range of average delay is the same for all the algorithms including ours when the sending rate is 2 packets/s. The maximum average delay of Pop-aware is 3500 ms that is 13.4% less than wGPD, 15.5% less than CaSMA and 22.82% less than S-R-1. The reason for high delay in S-R-1 is its consideration of only homogeneous delay rather than heterogeneous delay. However, our scheme transfers high loaded data packet first and use proper scheduling decision method to overcome this issue. The proper scheduling decision method solved the delay issue after considering the limited number of node's for transferring the data packets. As displayed in Fig. 7(b), the average throughput of a node is directly proportional to packet sending rate. Therefore, Pop-aware gained a higher advantage in comparison to other existing schemes for achieving the maximum average throughput after assigning the prioritization based on high degree centrality and *active service* is not yet available. Due to this fair allocation of resources, Pop-aware achieved maximum throughput. Generally, the performance of our proposed algorithm is better than other existing algorithms both for average delay and average throughput.

In Figs. 8(a) and (b), we present the evaluation of the proposed Pop-aware scheme with and without scheduling in order to show its significance and the standard deviation of average values are also illustrated in figures. The plots clearly show the distribution of data packets among nodes. ASNETs based nodes perform irregular distribution of data packets when scheduling algorithm is not applied. This is because of concentration of high data transfer through one intermediate node. Moreover, the other reason for bad performance without scheduling in ASNETs is due to the social property based communication between nodes. These social properties are helpful to reduce the congestion from the network after making proper scheduling decisions. As shown in Fig. 8(a) regarding Pop-aware with scheduling, almost 63.5% packet delivery ratio is achieved when the number of connections are 50 and there is no congestion throughout the network. However, the same plot with no scheduling scheme shows that the maximum rate of delivery is 41% when the numbers of connections are 50. Furthermore, Fig. 8(b) defines the average delay with maximum number of connections. It shows that the average delay using scheduling algorithm is 4500 ms and the result without applying scheduling algorithm is 8000 ms. This is due to the accurate decision of transferring data packets based on our purposed scheduling scheme. Overall, our proposed algorithm is superior in terms of the adopted evaluation metrics in comparison to the other algorithms.

## 7. Conclusions

In this paper, we have proposed Pop-aware, a congestion control scheme based on popularity-aware scheduling combined with a load estimation technique. This scheme provides higher priority to the flow of data packets that have high popularity level. To prevent dropping of prioritized node packets and congestion at an intermediate node, our method uses an efficient scheduling algorithm. Moreover, Pop-aware provides a fair dissemination of services among nodes and also a solution to the new incoming data packets. Pop-aware firstly calculates the load of an intermediate node when it considers that the node has become too congested. Secondly, it uses degree centrality social property for assigning the prioritization among data packets. The effectiveness of the proposed approach is demonstrated through extensive simulations. The results show that the proposed scheduling algorithm is significantly better than the existing algorithms for comparison. As compared to existing algorithm, Pop-aware resolved better overhead issue with efficient resource utilization and the throughput of our algorithm is higher due to scheduling of data packets that are based on degree centrality. In addition, our algorithm is also capable to minimize the delay in transmission and packet loss rate. Pop-aware efficiently distributes data of most popular node among nodes and reduces the overall congestion in the network.

In our future work, we plan to design popularity based data rate adjustment at the sender or receiver node. Our aim will be to provide reliability to the popular node after sending early acknowledgment packets to the popular sender node. The system will also reduce overhead on the network after delaying the acknowledgment packets. Pop-aware will need to integrate an algorithm that considers the mobility of nodes in the social groups or community, in order to further reduce the overhead incurred by the link's contention between sender and receiver nodes.

## Acknowledgment

This work is partially supported by the Fundamental Research Funds for the Central Universities (DUT15YQ112).